# Interface induced high temperature superconductivity in single unit-cell FeSe films on SrTiO$_3$

Qing-Yan Wang[1, 2, #], Zhi Li[2, #], Wen-Hao Zhang[1, #], Zuo-Cheng Zhang[1, #], Jin-Song Zhang[1], Wei Li[1], Hao Ding[1], Yun-Bo Ou[2], Peng Deng[1], Kai Chang[1], Jing Wen[1], Can-Li Song[1], Ke He[2], Jin-Feng Jia[1], Shuai-Hua Ji[1], Yayu Wang[1], Lili Wang[2], Xi Chen[1], Xucun Ma[2, *], and Qi-Kun Xue[1, *]

[1] State Key Lab of Low-Dimensional Quantum Physics, Department of Physics, Tsinghua University, Beijing 100084, China

[2] Institute of Physics, The Chinese Academy of Sciences, Beijing 100190, China

[#] Authors equally contributed to this work.

* Correspondence and request form materials should be addressed to Q.K.X. (qkxue@mail.tsinghua.edu.cn) or X.C.M. (xcma@iphy.ac.cn).

Searching for superconducting materials with high transition temperature (T$_C$) is one of the most exciting and challenging fields in physics and materials science. Although superconductivity has been discovered for more than 100 years, the copper oxides are so far the only materials with T$_C$ above 77 K, the liquid nitrogen boiling point[1,2]. Here we report an interface engineering method for dramatically raising the T$_C$ of superconducting films. We find that one unit-cell (UC) thick films of FeSe grown on SrTiO$_3$ (STO) substrates by molecular beam epitaxy (MBE) show signatures of superconducting transition above 50 K by transport measurement. A superconducting gap as large as 20 meV of the 1 UC films observed by scanning tunneling microcopy (STM) suggests that the superconductivity could occur above 77 K. The occurrence of superconductivity is further supported by the presence of superconducting vortices under magnetic field. Our work not only demonstrates a powerful way for finding new superconductors and for raising T$_C$, but also provides a well-defined platform for systematic study of the mechanism of unconventional superconductivity by using different superconducting materials and substrates.

The interface enhancement of electron-phonon coupling[3] and epitaxial strain[4] have been separately employed to increase the $T_C$ of superconductors. In this study we attempt to employ both effects by growing ultrathin films of superconducting β-phase FeSe[5] on STO (001) substrates, which have a dielectric constant of ε = 300. The system has a lattice mismatch of 1% (the lattice constant of bulk FeSe is smaller than of STO). Under usual preparation conditions, the STO surfaces always contain some Sr or Ti clusters and other defects. We develop a new technique named Se molecular beam etching to obtain atomically smooth STO surface (see Methods). The resulted surface morphology is shown in Fig. 1a, which is basically free of defects. An atomically sharp interface between FeSe and STO is thus expected. We choose the binary alloy FeSe[5], the simplest material among the recently discovered iron-based superconductors[6], as the first system to test our idea simply because the MBE growth conditions for stoichiometric and single crystalline FeSe films have been well established[7, 8].

FeSe grows on the Se-etched STO(001) via a typical layer-by-layer mode. Shown in Fig. 1b is a scanning tunnelling microscopy (STM) topographic image of the atomically flat surface after deposition of about one unit-cell (1 UC) thick FeSe film. One UC FeSe along the c-axis is made of a Se-Fe-Se triple layer and has a thickness of 0.55 nm on STO, as schematically shown in Fig. 1c. The zoom-in STM image in Fig. 1d reveals a perfectly ordered Se-terminated (001) lattice, the same to that of FeSe grown on graphene/SiC(0001)[7]. The in-plane lattice constant is 3.8 Å, suggesting a 1% tensile strain in the FeSe films (see Supplementary Fig. S1). High energy resolution scanning tunnelling spectroscopy (STS) measurement reveals a clear signature of superconductivity. Fig. 1e shows the tunnelling spectrum taken on the 1 UC FeSe at 4.2 K. The film exhibits an overall U-shaped conductance spectrum: a zero conductance region near the Fermi level ($E_F$) and an unusually large superconductive gap Δ = 20.1 meV defined by the distance between the two sharp peaks in Fig. 1e. This value is almost one order of magnitude larger than Δ ~2.2 meV for bulk FeSe ($T_C$ = 9.4 K[9]) measured using the same instrument[7,8]. The ratio of $2\Delta/k_B T_C$ is ~5.5 ($k_B$ is the Boltzmann constant) for bulk FeSe. If we assume the same superconducting mechanism held for both the free-standing and strained FeSe films, the gap of 1 UC FeSe will lead to a superconducting transition at ~80 K. Although such estimation is very rough, we expect that the transition temperature could very likely exceed 77 K-the liquid nitrogen boiling point. The FeSe/STO heterostructure is the first system we have tested with the above sample preparation method. Optimization of $T_C$ with improved FeSe/STO interface quality and with other heterostructure systems can be envisioned.

While our preliminary variable temperature tunnelling measurement shows additional evidence for superconductivity (Supplementary Fig. S2), the occurrence of superconductivity in the 1 UC films on STO is further confirmed by the presence of superconducting vortices under external magnetic field at 4.2 K. Figure 2a shows the zero bias conductance spectra mapping of a surface region shown in Fig. 2b, where a vortex is clearly observed. Figure 2c displays a series of tunnelling spectra taken at the points indicated by the dots in Fig. 2a. Towards the vortex center, the coherence peaks are gradually suppressed while the gap size remains unchanged.

We find that the second UC and thicker films do not superconduct at all, and the observed superconductivity behaviour is limited to the very first unit cell of the film above the interface. (For proximity effect between 1 UC and 2 UC films, see Supplementary Fig. S3.) Shown in Fig. 1f is a tunnelling spectrum taken on the 2 UC thick films. There is no superconducting gap and its electronic structure near $E_F$ is characterised by a semiconductor-like behaviour. The feature is in sharp contrast to the free-standing FeSe films grown on graphene/SiC(0001) where the $T_C$ increases almost linearly with increasing film thickness[8]. The difference documents a significant role of the FeSe/STO interface in the observed superconductivity.

It is difficult to directly measure the superconducting properties of the above mentioned 1 UC FeSe films by transport measurement. The main reason for this difficulty is that the STO surface after Se beam etching at 950 ºC becomes very conductive with resistivity in the order of $10^{-4}$ Ω·cm. To carry out transport measurement, we have to use the insulating STO(001) substrates that were only treated by $O_2$ in a tube furnace (Methods). A film of the 5 UC FeSe was covered with a 20 nm thick amorphous Si protection layer for *ex situ* transport measurement (Supplementary Fig. S4 and S5). As shown in Fig. 3a, the temperature dependent resistance (R-T) clearly reveals the occurrence of superconducting transition with an onset temperature of 53 K. This value is the highest among more than 30 films grown under the same condition. Typically, a value of ~40 K is obtained. The superconducting transition is suppressed by magnetic field (see the upper insert in Fig. 3a), a typical characteristic of superconductors. To correlate the $T_C$ with gap $\Delta$, we carried out low temperature tunneling spectroscopy measurement. Figure 3b shows the tunnelling spectrum of 1 UC film grown on the insulating STO using the same condition. A gap of ~10 meV is clearly observed. Using the same BCS ratio (5.5) mentioned above, we obtain $T_C$ = 42 K, which agrees with the transport experiment. Since the 2 UC and thicker films are non-superconducting (see Fig. 3c), the transport measurement shown in Fig. 3a should only reflect the superconductivity of the first UC FeSe. In order to determine the $T_C$ associated with $\Delta$ = 20.1 meV directly by transport, preparation of atomically flat insulating STO, which may be done with *in situ* MBE or pulse laser deposition without surface treatment, or using other insulating substrates is necessary. We leave this for future experiments.

While the mechanism for this high Tc superconductivity is not completely clear for the time being, we argue that the interface plays the dominant role. According to our recent study on ultrathin FeSe films (from 1 UC to 8 UC) grown on graphene/SiC (its dielectric constant $\varepsilon$ < 10), the upper limit of $T_C$ for unstrained 1 UC FeSe is 2 K[8]. For bulk FeSe, by applying external pressure $T_C$ can increase by four times (from 9.4 K to 36.7 K) due to lattice compression [10]. Assuming a similar enhancement effect by the epitaxial strain here and taking a simplest estimation, a $T_C$=8 to 10 K for 1 UC FeSe on STO would be expected. However, this effect is too weak to account for the observed value. One must consider another interface effect, the interface enhanced electron-phonon coupling[3,11] at the FeSe/TiO interface, as demonstrated in monolayer Pb and In films on Si(111) with a very similar structure[12,13]. In the present case, the effect may be further promoted by the polaronic effect associated with the high dielectric constant of STO. Further investigation is needed to elucidate the mechanism underlying the observed superconductivity.

Before closing, we would like to point out some implications of our study. (1) In principle, the present method can be applied to any existing superconductor material on any substrate as long as atomically sharp and strongly bonded interface can be formed experimentally. Taking K-doped FeSe[14] as an example, it is recently demonstrated that high quality films of K-doped FeSe can be prepared on graphene by MBE[15]. If high dielectric constant substrates such as $BaTiO_3$ and $SrTiO_3$ can be used to achieve similar interface effect, one may expect much higher $T_C$. Therefore, our study points out a straightforward direction to find superconductors with very high $T_C$. (2) We note that there is a remarkable resemblance in the bonding configuration between FeSe-TiO at FeSe/STO interface and that of cuprates superconductors, for example CuO-SrO in BSCCO, and that of iron-pnictide superconductors, for example FeAs-LaO in LaOFeAs. From this point of view, the results presented in this work provide crucial clue for revealing the secret of unconventional superconductivity: the high $T_C$ of the layered cuprates may very likely result from a single unit cell of the material[16]. By systematically varying the superconducting material and substrate with different dielectric and lattice constants, one can pin down the effect responsible for the gluing mechanism of Cooper pairs. (3) By depositing dielectric gate material on top of the epitaxial superconducting films, which can easily be done with the present method, further enhancement in $T_C$ by electrical field effect may be achieved. (4) Because our approach to raising $T_C$ is based on high quality ultrathin films with atomic-layer perfection over macroscopic scale by standard MBE technique, one can easily employ it to develop superconductor electronics and other applications.

**Methods**

The Nb doped (0.5wt%) and (001) orientated single crystal STO (Shinkosha) was used in the MBE and low temperature STM/STS combined system (Unisoku). Before FeSe thin film growth, the substrate was degassed at 600 °C for 3 hours and then heated to 950°C under the Se flux for 30 min. High quality of the sample surface after the treatment was demonstrated by the sharp RHEED patterns as shown in Fig. S1. The FeSe films were grown by co-evaporating Fe (99.995%) and Se (99.9999%) from standard Knudsen cells with a flux ratio of approximately 1:10 at 450 °C. The Fe flux is approximately 0.06 ML/min. The FeSe thin films was gradually annealed to 550 °C by several steps. A polycrystalline PtIr tip was used and the STS was acquired using lock-in technique with a bias modulation of 1mV at 931 Hz . The STM images were processed using WSxM software.

For the transport experiment, the insulating single crystal STO (001) (2 mm × 10 mm × 0.5 mm) was used as the substrate for the growth of FeSe films. The STO(001) was first pretreated by a two-step chemical etching and thermal annealing method in order to obtain a specific $TiO_2$-terminated surface. In the first step, it was soaked in hot (350 K) deionized water for 45 min and immersed in 10% HCl solution for 45 min to dissolve the outmost SrO layer. In the second step, STO was annealed at 900 °C for 2 h in a tube furnace with oxygen-argon (3:1) mixed gas. After the pretreatment, STO was transferred into UHV MBE chamber (Omicron) and further annealed at 580 °C for 3 h. Through these treatments, the STO(001) surface becomes flat and shows typical step-plus-terrace structure. An amorphous Si film was deposited on the FeSe films at 150 K as a protection layer before they were transferred out of the growth chamber. The nominal thickness of the amorphous Si film is 20 nm (see Supplementary Fig. S4). The transport measurements were performed using the standard four-probe ac lock-in method.

**Acknowledgements.** We acknowledge stimulating discussion with Xincheng Xie and Fu-Chun Zhang. The work was financially supported by National Science Foundation and Ministry of Science and Technology of China.

**Author Contributions.** Q.K.X., X.C.M. and X.C. conceived the study. L.L.W. and X.C.M. designed MBE growth and room temperature STM experiment, X.C. and Q.K.X. designed MBE growth and low temperature STM/STS experiment, and Y.Y. W. designed the transport experiment. Q.Y.W, W.H.Z., Y.B.O., L.L.W. and K.H. carried out MBE growth and room temperature STM experiment. Z.L., S.H.J., W.L., H.D. P.D., K.C., and J.W. carried out MBE growth and low temperature STM/STS experiment. Z.C.Z, J.S.Z., W.H.Z and Y.Y.W. carried out transport measurements. C.L.S., J.F.J., S.H.J., L.L.W., Y.Y.W., X.C., X.C.M. and Q.K.X. assisted in the experiments and data analyses. Q.K.X. wrote the paper with contributions from S.H.J., L.L.W., Y.Y.W., X.C. and X.C.M. All authors discussed the results and commented on the manuscript.

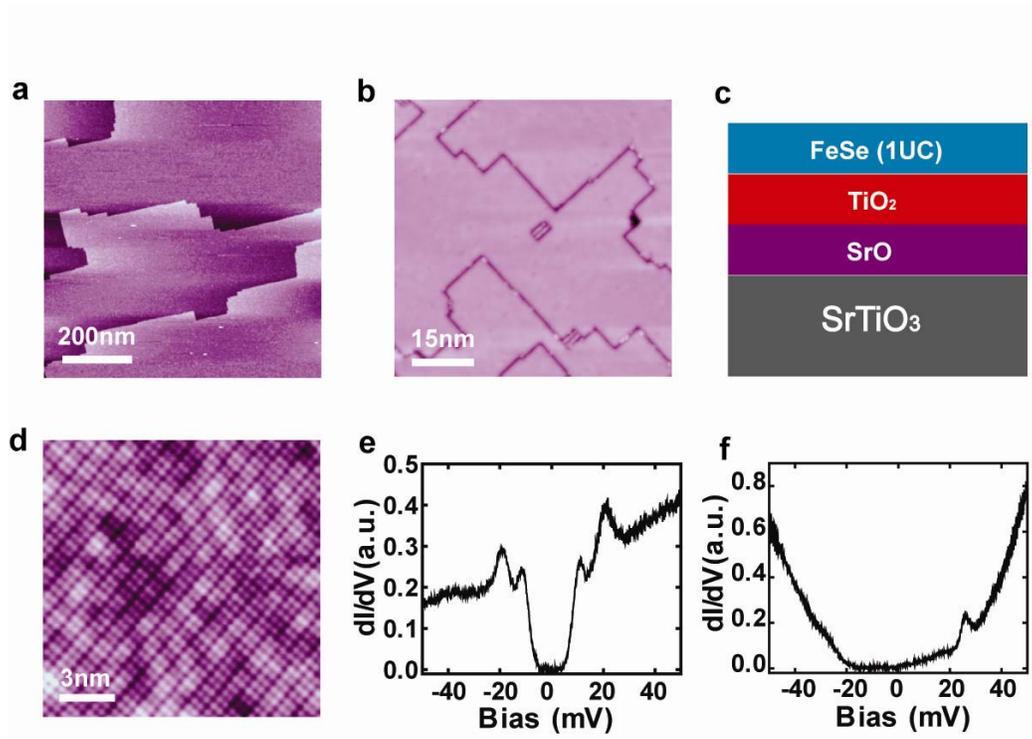

**Figure 1| Interface induced superconductivity of 1 UC thick FeSe films on STO(001).** **a,** STM topography (image size 800 nm × 800 nm, sample bias $V_S$ = 0.3 V, tunneling current $I_t$ = 24 pA) of STO(001) surface annealed at 950 °C under Se molecular flux in UHV MBE chamber. **b,** STM topography (67 nm × 67 nm) of 1 UC thick FeSe film on STO(001). Grain boundaries appear as trenches along <100> or <010> direction. $V_S$ = 3.1 V, $I_t$ = 29 pA. **c,** Schematic structure (side-view) of the FeSe films on STO substrate along the c-axis. Some of O atoms in $TiO_2$ plane might be substituted by Se atoms. **d,** Atomic resolved STM topography (12.8 nm × 12.8 nm) showing the Se terminated FeSe (001) lattice. $V_S$ = 0.6 V, $I_t$ = 51 pA. **e,** Tunneling spectrum taken on the 1 UC thick FeSe film on STO(001) at 4.2 K revealing the appearance of superconducting gap. Four pronounced superconducting coherence peaks appear at ±20.1 mV and ±9 mV, respectively. The zero differential conductance from −5 mV to 5 mV implies an s-wave-like (U-shaped) gap. **f,** Tunneling spectrum taken on the 2 UC thick FeSe films, which reveals a semiconductor-like (non-superconductive) behaviour. High resolution STM image (Fig. S3b) indicates that the in-plane lattice constants of the 1 UC and 2 UC films are the same.

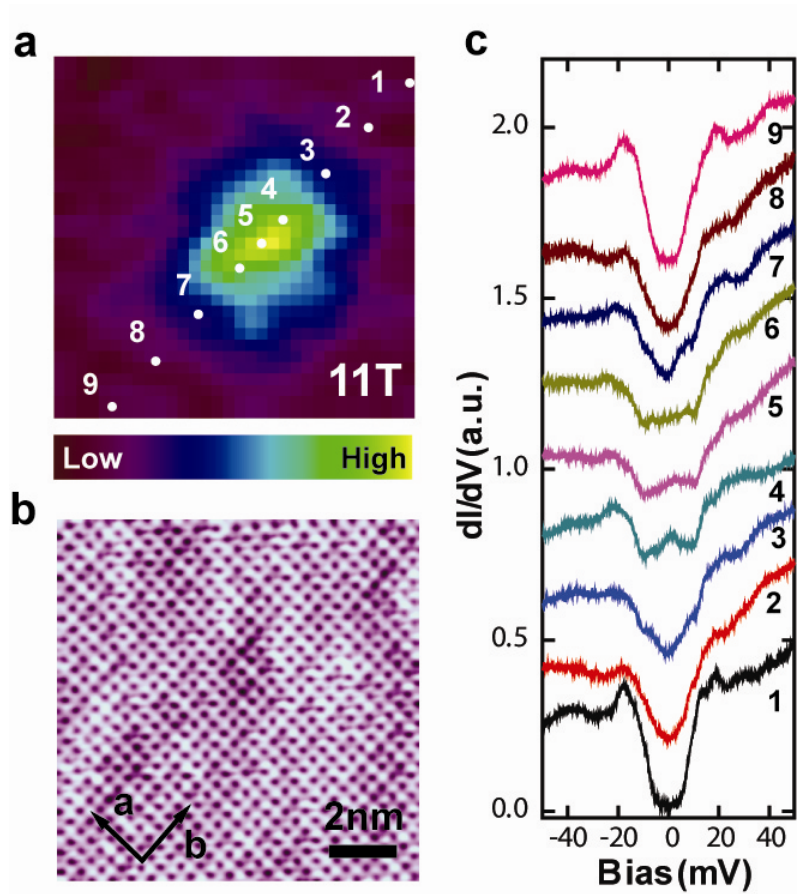

**Figure 2| Vortex of 1 UC FeSe film on STO. a,** Zero bias differential conductance mapping of the vortex state under magnetic field (11 Te) at 4.2 K. **b,** Simultaneously recorded STM topography (10.6 nm × 10.6 nm) of the mapping area shown in **a**. $V_S$ = 50 mV, $I_t$ = 52 pA. **c,** The scanning tunneling spectra on and near the vortex core. The locations where the spectra were taken were indicated by the white points marked in **a**. Near the vortex core center (points 4, 5 and 6), the superconducting coherence peaks at ~±20 meV disappear and bound states at the Fermi level appear. At different locations, there is no change in the superconducting gap size.

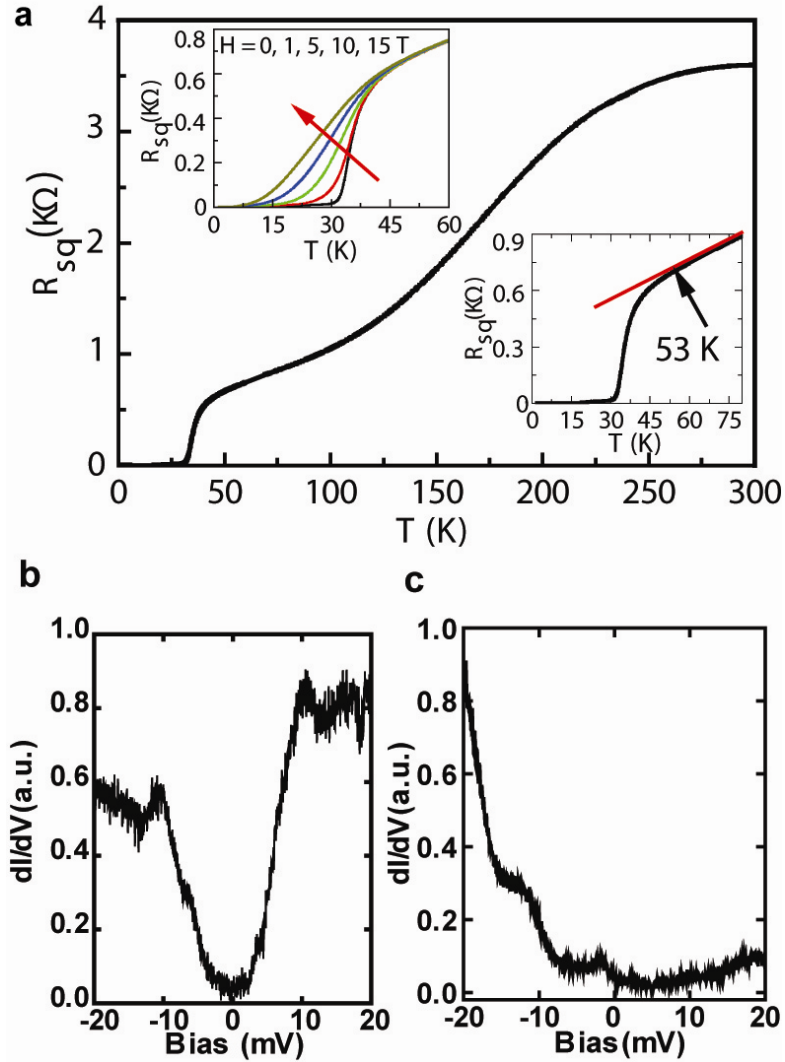

**Figure 3| The transport and STS measurements of the 1 UC thick FeSe films on insulating STO(001) surface. a,** Temperature dependence of square resistivity ($R_{sq}$) from 0 to 300 K. Upper insert: $R_{sq}$-T curves at various magnetic fields along c-axis, lower insert: $R_{sq}$-T curve from 0 to 80 K. **b,** The *dI/dV* spectrum of the 1 UC thick FeSe film on insulating STO(001) surface at 0.4 K ($V_S$ = 25 mV, $I_t$ = 99 pA). The gap as measured by two coherence peaks is ~10 meV. **c,** The *dI/dV* spectrum of the 2 UC thick FeSe film on insulating STO(001) surface at 4.2 K ($V_S$ = 25 mV, $I_t$ = 47 pA).

# Supplementary Information for

# Interface induced high temperature superconductivity in single unit-cell FeSe films on SrTiO$_3$

Qing-Yan Wang[1,2,#], Zhi Li[2,#], Wen-Hao Zhang[1,#], Zuo-Cheng Zhang[1,#], Jin-Song Zhang[1], Wei Li[1], Hao Ding[1], Yun-Bo Ou[2], Peng Deng[1], Kai Chang[1], Jing Wen[1], Can-Li Song[1], Ke He[2], Jin-Feng Jia[1], Shuai-Hua Ji[1], Yayu Wang[1], Lili Wang[2], Xi Chen[1], Xucun Ma[2,*], and Qi-Kun Xue[1,*]

[1] State Key Lab of Low-Dimensional Quantum Physics, Department of Physics, Tsinghua University, Beijing 100084, China

[2] Institute of Physics, The Chinese Academy of Sciences, Beijing 100190, China

[#] Authors equally contributed to this work.

* Corresponding authors: Q.K.X. (qkxue@mail.tsinghua.edu.cn) or X.C.M. (xcma@iphy.ac.cn).

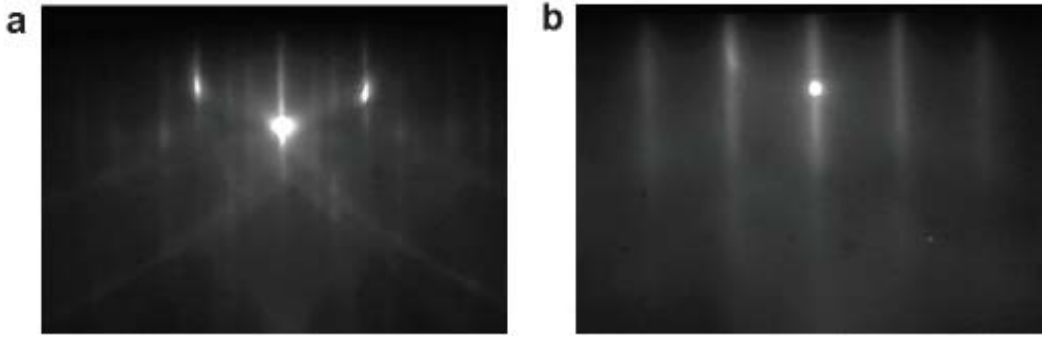

**Fig. S1. Reflection high electron energy diffraction (RHEED) patterns of the clean STO(001) surface and 1 UC FeSe on STO(001). a,** The RHEED pattern of the STO(001) substrate that was treated under Se flux at 950 °C. The in-plane lattice constant is 3.91 Å. The Kikuchi lines and sharp pattern indicate the atomically flat nature of the STO(001) surfaces prepared by our method. **b,** The RHEED pattern of 1 UC FeSe deposited on STO(001) surface shown in **a**, which shows that the in-plane lattice constant is 3.80 Å. According to the RHEED observation, the 1 UC FeSe films are expanded by 1% as compared to bulk FeSe.

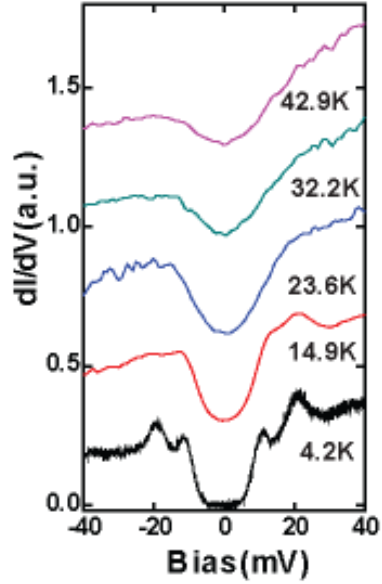

**Fig. S2. *dI/dV* tunnelling spectra of 1 UC FeSe on the Se-etched STO(001) surface at different temperatures.** From 4.2 K to 23.6 K, there is no obvious change in the gap size although the coherence peaks gradually fade out, which clearly reveals the robust superconductivity of the 1 UC films. In spite of the thermal drift, the superconducting gap is still clearly visible at 42.9 K. Measurement at higher temperatures (77 K) was attempted. However, due to very large thermal drifting in our STM system, no reliable spectra could be obtained.

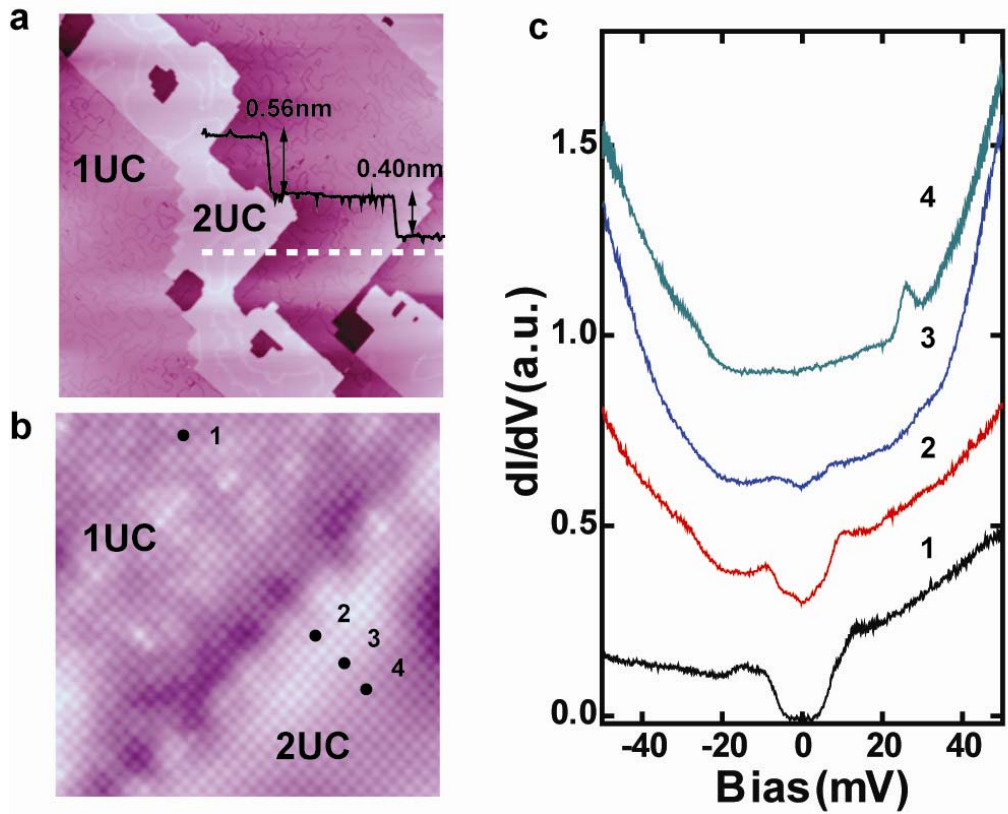

**Fig. S3. The superconducting proximity effect around boundary between 1 UC and 2 UC FeSe thin films. a**, The STM image (500 nm × 500 nm, $V_S$ = 2.75V, $I_t$ = 27 pA) of the FeSe film with the thickness more than one unit cell. The growth of the 2 UC FeSe starts from the step edges as shown in **a**. The dashed-line shows the topography profile across two steps. The 0.4 nm height step is the substrate step of STO and the 0.56 nm step is the single UC height between 1 UC FeSe film and 2 UC FeSe film. **b**, The STM image (13.2 nm × 13.2 nm, $V_S$ = 0.1 V, $I_t$ = 28 pA) of the boundary between 1 UC and 2 UC FeSe. **c**, The *dI/dV* spectra near the boundary. The spectra were taken at the locations 1, 2, 3 and 4 shown in **b**. The superconducting gap gradually disappears as the scanning tip moves into the 2 UC film region. It indicates a very short in-plane coherence length (around 4 nm).

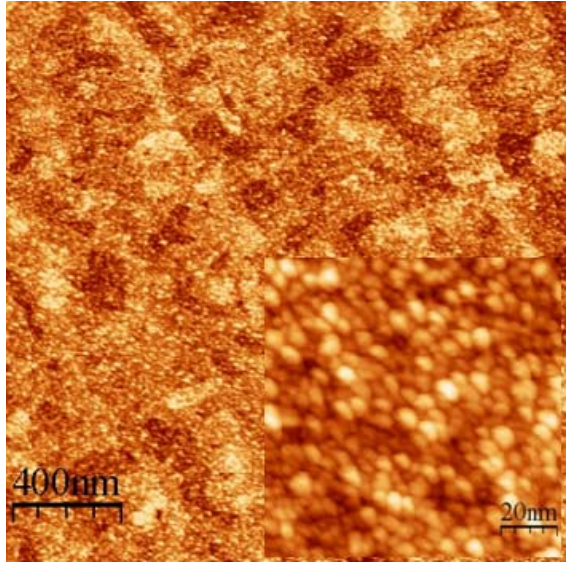

**Fig. S4. STM topography after deposition of a 20 nm thick Si capping layer on FeSe film.** The insert is a zoom-in image. The overall morphology is very similar to that of the grown FeSe film. The Si capping layer turns out to be a very effective protection layer preventing the FeSe film from ambient oxidation during *ex situ* transport.

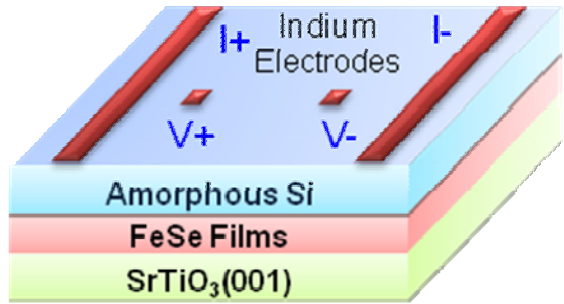

**Fig. S5. Schematic setup for transport measurement.** Since the STO substrate without Se beam treatment is insulating, the sandwiched FeSe films between the STO substrate and amorphous Si capping layer forms an ideal 2D electronic system for transport study. Electrical contacts to the FeSe film through the protection layer were achieved by pressing on the indium electrodes.